\documentstyle[aps,prc,preprint]{revtex}

\begin{document}
\draft

\title{\vspace*{-70pt}{\normalsize \hfill nucl-th/9505035}\\[-12pt]
{\normalsize \hfill KSUCNR-005-95}\\[-12pt]
{\normalsize \hfill McGill/95-26}\\[-12pt]
{\normalsize \hfill revised, November 8, 1995}\\[-12pt]
Electron-positron pairs from thermal resonances
in ultrarelativistic nuclear collisions}
\author{John J. Neumann,$^a$\thanks{Electronic mail (internet):
neumann@scorpio.kent.edu.}\ David Seibert,$^b$\thanks{Electronic
mail (internet): seibert@hep.physics.mcgill.ca.}\ and George
Fai$^a$\thanks{Electronic mail (internet): fai@ksuvxd.kent.edu.}}
\address{$^a$Center for Nuclear Research,
Department of Physics, Kent State University, Kent, OH 44242\\
$^b$Department of Physics, McGill University, Montr\'eal, QC H3A 2T8,
Canada}
\maketitle

\begin{abstract}
We use a boost-invariant one-dimensional (cylindrically symmetric)
fluid dynamics code to calculate $e^+e^-$ production from $\rho^0$
and $\omega$ decay in the central rapidity region of a central S+Au
collision at $\sqrt{s}=20$ GeV/nucleon.  We use equations of state
with a first-order phase transition between a massless pion gas and
quark gluon plasma, with transition temperatures in the range
$150-200$ MeV.  The production cross section at the $\rho$ mass
loosely constrains the transition and freeze-out temperatures,
and
we find that the $m_T$
spectrum is a good thermometer for sufficiently high $T_c$.
\end{abstract}

One of the most important problems in nuclear physics today is the study
of the equation of state (EOS) of high-temperature and high-density
hadronic matter; the former is necessary to understand the first few
microseconds of the big bang, and the latter for elucidating the physics
of neutron stars and similar dense astrophysical objects.  In order to
investigate the nature of the
transition from normal ($T=0$) hadronic matter to deconfined quark gluon
plasma (QGP), or whatever the high temperature phase may be, experiments
are being performed at Brookhaven's Alternating Gradient Synchrotron
(AGS) and CERN's Superconducting Proton Synchrotron (SPS), and future
experiments are being planned for Brookhaven's Relativistic Heavy Ion
Collider (RHIC) and CERN's Large Hadronic Collider (LHC).  In these
experiments, the only particles expected to emerge from the early part of
the collision (and thus the only probes of the high-temperature EOS) are
photons and leptons, which therefore may provide the most important
information obtained from these collisions.

Predicted $e^+e^-$ transverse mass, $m_T$, spectra from resonance decays
in ultra-relativistic nuclear collisions are sensitive to the hadronic
transition temperature, $T_c$, when transverse expansion is
neglected\cite{rho,phi,phi2}.  Thus, measurements of these spectra may
place strong constraints on $T_c$, as well as other parameters of the
hadronic EOS\cite{phi2}.  However, the photon transverse momentum spectrum
was also predicted to be sensitive to $T_c$\cite{phth}, but the
sensitivity disappeared when transverse expansion was included\cite{jn}.
The next natural step is to study $e^+e^-$ transverse mass spectra with
a model including transverse expansion. This is the subject of the present
paper. Since transverse expansion may increase the contribution
of mesons decaying after freeze-out, a careful treatment of freeze-out
is necessary. We find that the $m_T$ spectra are in fact
dominated by the freeze-out contribution for sufficiently high $m_T$,
but can still be fitted with apparent temperatures correlated to $T_c$.
This may provide a method to determine $T_c$ experimentally, even though
the underlying physics is more complicated than originally envisioned
in the treatment of lepton pairs without transverse expansion\cite{rho}.
We show that the correlation between $T_c$ and the $m_T$
spectrum of dileptons from $\rho^0$ and $\omega$ decay persists in the
presence of transverse expansion,
 and thus that a determination of the
transverse mass spectrum of dileptons in the $\rho^0-\omega$ peak would
yield useful constraints on $T_c$.

Here we adopt the position that in strongly-interacting matter at
sufficiently high temperature and/or density a transition takes place
to a high-energy phase. In the present paper, we are interested in
the possibility of extracting the transition temperature assuming a
first-order phase transition.
We assume a boost-invariant longitudinal
expansion as discussed by Bjorken\cite{Bj}, coupled to a cylindrically
symmetric transverse expansion.  For the initial evolution, we use a
simple model for production of longitudinally free-streaming hot
matter\cite{s0}.  After the hot matter has thermally equilibrated,
we use thermal equilibrium fluid-dynamical evolution, but consider
deviations from chemical equilibrium in the high-temperature phase by
allowing the quark and antiquark densities to be a (fixed) fraction, $x$,
of their equilibrium values.  When particle mean free paths become
comparable to the radius of the cylinder of hot matter, which we take to
occur at some freeze-out temperature, $T_{fo}$, we
assume that the particles crossing the freeze-out surface $T=T_{fo}$
stream freely until they either decay or reach the detectors.
We then calculate the $e^+e^-$
production rate from $\rho^0$ and $\omega$ decays; these are not separated,
as the masses are almost degenerate and the experimental resolution is
typically not good enough to resolve the individual peaks.
We investigate the sensitivity to different assumptions about
the initial temperature, freeze-out temperature, and
quark fraction, and compare production rates to the preliminary
NA45 data\cite{ull}.
We use standard high-energy conventions, $c=\hbar=k_B=1$.

Here we describe the initial conditions and  assumptions about the
EOS; the details of the fluid-dynamical calculation can be found
elsewhere\cite{blol,jn}.  We consider a central collision of two large
nuclei at SPS energy ($\sqrt{s}=20$ GeV/nucleon).  For such high
collision energies we expect approximate longitudinal boost
invariance\cite{Bj}, so the behavior of the produced matter at
different rapidities is the same in the longitudinally comoving frame
for fixed proper time $\tau=\sqrt{t^2-z^2}$, where $z$ is the
distance along the beam axis.  At $\tau=0$ the colliding nuclei reach
the point of maximum overlap and are assumed to form a longitudinally
expanding pancake.  The hot matter has thermalized at $\tau=\tau_0$
($\approx$~0.2 fm/$c$)\cite{shur,gkce}, when a
cylindrically-symmetrical transverse expansion begins, coupled to the
longitudinal expansion.

{}From $\tau=0$ until the transverse expansion starts at $\tau=\tau_0$,
we assume a boost-invariant cylinder of radius $R_<$ (the radius of the
smaller nucleus), filled uniformly with QGP at
temperature $T=T_0$.  This is approximately compatible with the initial
entropy density for short times\cite{s0}.  We determine $T_0$ by assuming
entropy conservation for $\tau > \tau_0$, hence
\begin{equation}
s(T_0)=\frac{3.6 \, dN_\pi/dy}{\pi \, R_<^2 \, \tau_0},
\end{equation}
where $s$ is the entropy density, with total (charged plus neutral)
multiplicity density $dN_\pi/dy$.

The equations of state (EOS's) that we use here are of the form
\begin{equation}
{\renewcommand{\arraycolsep}{2pt}
\begin{array}{lrclrcl}
T<T_c:\quad &
 & e & = \displaystyle \frac{\pi^2}{10} T^4, \quad &
 & P & = \displaystyle \frac{e}{3}, \quad \\[12pt]
T=T_c:\quad &
 \quad \displaystyle \frac{\pi^2}{10} T_c^4 \leq & e &
  \leq \displaystyle \frac{\pi^2}{30} g_q T_c^4 +B, \quad &
 \quad \displaystyle \frac{\pi^2}{30} T_c^4 = & P &
  = \displaystyle \frac{\pi^2}{90} g_q T_c^4 -B, \quad \\[12pt]
T>T_c:\quad &
 & e & = \displaystyle \frac{\pi^2}{30} g_q T^4 +B, \quad &
 & P & = \displaystyle \frac{\pi^2}{90} g_q T^4 -B, \quad
\end{array}
}
\label{EOS}
\end{equation}\\
where $g_q$ is the number of massless degrees of freedom
in the deconfined phase.
We treat only the case of zero baryon density, so the entropy density is
$s =(e+P)/T$
independent of the phase of the matter.  Below $T_c$, the EOS is that of
a massless pion gas.  Because recent calculations have
predicted that the quarks may reach only a
fraction of their equilibrium number by the beginning of
transverse expansion\cite{shur,gkce}, we take
$g_q =16+21x$, where $x$ is a parameter
that we vary between 0 and 1 to simulate
 the effect of reducing the quark density in the QGP
 below the equilibrium value
($x$=1 is equilibrium for two flavors of massless quarks).
The vacuum energy density in the deconfined phase, $B$, is related to
the transition temperature, $T_c$, by requiring equal pressures in
the deconfined and hadronic phases at $T=T_c$ so that
$B=\pi^2 (g_q-3)T_c^4/90$.
The value of $B$ we use away from equilibrium
is calculated by assuming
$x=1$ and supplying an equilibrium transition temperature $T_c^\prime$.
It is this $T_c^\prime$ that is actually used on the graphs and is
useful for comparison purposes.

We calculate the central rapidity region $m_T$ distribution from both
interacting and frozen-out hot matter.  The contribution from the
interacting hot matter is
\begin{eqnarray}
\left. \frac {d^2N_{e^+e^-}^{(\rho^0+\omega)}} {m_T dm_T dy} \right|_{y=0}
= \int d\eta \int d\tau \, \tau \int dr \, 2\pi r
\int_0^{\pi} d\theta \, \sin\theta \int_0^{2\pi} d\phi
\int_0^{\infty} dp \, p^2 \, \frac {dR} {d^3p}  \label{bulkeq} \\[12pt]
\times
\delta\left( \frac 1 2 m_T^2 - \frac 1 2 m_T^{\prime 2} \right) \,
\delta\left( \eta +\tanh^{-1} \left[ \frac {p \cos\theta}
{\gamma \left(\sqrt{p^2+m^2}  +  p \, v\sin\theta\cos\phi \right)}
\right] \right) \, \Theta (T-T_{fo}). \nonumber
\end{eqnarray}
Here $m_T=\sqrt{m^2+p_T^2}$,
$R$ is the $e^+e^-$ production rate per unit four-volume in the
fluid frame, $v$ is the transverse velocity of the fluid
in the cell characterized by proper time $\tau$, space-time rapidity $\eta$
and radial position $r$ (measured in the frame
moving with transverse velocity zero and longitudinal velocity
$\tanh\eta$ in the lab), and $\gamma=(1-v^2)^{-1/2}$.
The $e^+e^-$ production rate from $\rho^0$ and $\omega$ mesons in
thermally and chemically equilibrated hadron gas is
\begin{equation}
\frac {dR} {d^3p} ~=~  \frac {m (g_{\rho^0}\Gamma_{\rho^0 \rightarrow e^+e^-}
+g_{\omega}\Gamma_{\omega \rightarrow e^+e^-})}
{E (2\pi)^3} \left( e^{E/T}-1 \right)^{-1}, \label{ERse}
\end{equation}
where $g_{\rho^0}=g_{\omega}=3$ are the degeneracies of
the $\rho^0$ and $\omega$ mesons, and the partial widths are
 $\Gamma_{\rho^0 \rightarrow e^+e^-}=6.77$ keV and
$\Gamma_{\omega \rightarrow e^+e^-}=0.60$ keV.
$E$ is the energy  measured in the fluid frame,
and we use the average mass of the $\rho$ and $\omega$, $m=0.775$ GeV.
No mesons exist in the QGP, and hence the resonant $e^+e^-$ production
rate is zero for that part of the fluid.

Mesons which pass through the freeze-out surface $T=T_{fo}$ are no
longer subject to fluid-dynamic flow, as their mean free paths
are so long that they do not interact with the surrounding matter.
As these frozen-out mesons free-stream toward the detectors they
decay to $e^+e^-$ pairs as before, but the number of mesons decreases
exponentially in time, with time constant equal to the free space
meson decay time.  The contribution from the decay of frozen-out
$\rho^0$ and $\omega$ mesons is thus

\begin{eqnarray}
\left. \frac {d^2N_{e^+e^-}^{(\rho+\omega)}}{m_Tdm_Tdy} \right|_{y=0}=
\left(\frac {g_{\rho^0}\Gamma_{\rho^0 \rightarrow e^+e^-}}
{\Gamma_{\rho}}+
\frac {g_{\omega}\Gamma_{\omega
\rightarrow e^+e^-}}{\Gamma_{\omega}} \right)
\int d\tau \, \tau \, 2\pi \, r(\tau) \int_{-1}^{1} d(\cos \theta)
\int_0^{2\pi} d\phi \label{freeze} \\[12pt]
\times \int_0^{\infty} dp \, p^2 \gamma
(2\pi)^{-3}\left( e^{E/T}-1 \right)^{-1}
\left( v_x - \frac {dr}{d\tau} \right)
\Theta \left( v_x - \frac {dr}{d\tau} \right)
\delta(\frac 1 2 m_T^2 - \frac 1 2 m_T^{\prime 2}) \nonumber
\end{eqnarray}

where
\begin{center}
\begin{equation}
v_x = \frac{p \sin\theta \cos \phi +v\sqrt{p^2+m^2}}
{\sqrt{p^2+m^2}+ p \, v  \sin\theta \cos\phi}
\end{equation}
\end{center}
is the component of the meson velocity perpendicular to the
freeze-out surface,
and $\Gamma_{\rho}=151$ MeV and $\Gamma_{\omega}=8.43$ MeV
are the total widths for the $\rho$ and $\omega$.
$dr/d\tau$ is the radial velocity of the freeze-out surface
and $r(\tau)$ is its location.
The $\Theta$ function ensures that only mesons going through the
freeze-out surface in the outward direction are counted.

Finally, we extract the fit temperature, $T_{fit}$, by performing a
least-squares fit to our calculated spectrum (equilibrium plus
freeze-out) in the  window $1.155\le m_T \le 1.755$ GeV,
which is dominated by freeze-out.
The part of the spectrum that is dominated by the equilibrium
contribution is much smaller, typically $0.8\le m_T\le
0.9$ GeV. Unfortunately this is of the order of a single
experimental bin, so we choose instead to focus on the
freeze-out contribution, in anticipation of fitting experimental data
with both equilibrium and freeze-out contributions.
Furthermore, a measurement of an $m_T$ spectrum from the $\rho-\omega$
peak would look at a
finite range of $m$, so we filter the simulated data through an invariant mass
bin of size $0.475<m<1.075$ GeV,
assuming a Breit-Wigner distribution for each species.
For our fitting function, we take eq.~(\ref{bulkeq}) with $v=0$
(i.e.\ ignoring transverse expansion), resulting in the formula\cite{rho}
\begin{equation}\label{tfit}
\frac{d^2N_{e^+e^-}^{(\rho+\omega)}}{m_T dm_T dy} ~\sim~
\left( \frac{m_T}{T_{fit}} \right)^{1/2} \exp \left(
-\frac{m_T}{T_{fit}} + 0.4 ~\frac{T_{fit}}{m_T} \right).
\end{equation}

Our ``standard'' parameter set is $\tau_0$= 0.2 fm/$c$, $x$=1,
$T_{fo}=120$ MeV, and $150 < T_c < 200$ MeV; if different values of any
of the parameters are given, all the others are held at the standard
values.  We use $dN_\pi/dy=188$ for central S+Au collisions at SPS
energy (the NA45 experimenters estimate $dN_{ch}/d\eta=125$ for their
S+Au collision sample, and we assume isospin symmetry).  Our standard
value for the equilibration time, $\tau_0$= 0.2 fm/$c$, implies an
initial temperature $T_0=327$ MeV.
Fig.~1 shows the equilibrium, freeze-out and total
contributions for the standard set with $T_c$=170 MeV.

In the absence of transverse expansion, $T_{fit}$ is a monotonically
increasing function, and comparable to
$T_c$, so one can infer $T_c$ given $T_{fit}$
from the measured $e^+e^-$ spectrum\cite{rho}.
In Fig.~2, we vary $T_c$ from 150 to 200 MeV and calculate the
resulting values of $T_{fit}$ while including transverse expansion.
We find that $T_{fit}$ is higher than in
the case of longitudinal-only expansion,
but still monotonic (except where $T_{fo}$ approaches $T_c$).
$T_{fit}$ is 10-20 MeV lower for $\tau_0$=1 fm ($T_0$=192 MeV), which
we attribute to the fact that transverse expansion develops relatively
little in this case.

Note that freeze-out is treated dynamically in the present
calculation in contrast to an earlier investigation \cite{rho}
where transverse expansion was neglected and thus freeze-out was
approximated as occurring at a fixed proper time.
We consider a moving freeze-out surface here, and the yield
from mesons crossing this surface is integrated over the history
of the collision.
This contribution is enhanced by transverse expansion; the larger
the temperature difference $T_c-T_{fo}$, the stronger the enhancement.
We find that increasing $T_{fo}$ to 140 MeV from its standard
value decreases $T_{fit}$.
For larger $T_{fo}$ the hadronic matter near the
freeze-out surface
expands outward at a lower velocity, so that there is less
contribution to $m_T$ from the fluid motion.
Thus $T_{fit}$, which depends strongly on the fluid motion, is lowered.

The curve for $x=0.7$ shows that $T_{fit}$ is about 15 MeV
higher than for $x=1$.  The increase
occurs because the temperature of the mixed phase, which makes a large
contribution to the $e^+e^-$ spectrum, is raised when $x$ is lowered
and $B$ is held constant.

Our overall production
at the $\rho^0-\omega$ peak is up to a factor 3 below the value
measured by the NA45 experiment.
We take the value measured in the
experimental bin $0.7<m<0.8$ GeV (which contains the
$\rho^0 - \omega$ peak),
assuming that the signal is all from meson
decay, obtaining $d^2N_{e^+e^-}^{(\rho^0+\omega)}/dm
d\eta=1.7 \times 10^{-3}$ GeV$^{-1}$.
We calculate the same quantity for our simulated data from the
standard calculation with $T_c$=150 MeV, including
the experimental acceptance by counting only $e^+e^-$
pairs whose members both have $2.1 < \eta < 2.65$, and integrating
over a Breit-Wigner distribution for each species,
giving us $d^2N_{e^+e^-}^{(\rho^0+\omega)}/dm d\eta=
5.9 \times
10^{-4}$ GeV$^{-1}$ for our standard
parameter set.
Increasing $T_c$ to 200 MeV makes $d^2N_{e^+e^-}^{(\rho^0+\omega)}
/dm d\eta = 1.2 \times 10^{-3}$ GeV$^{-1}$, still below
the preliminary NA45 data.
Additional pair-producing processes not included here
will increase the cross-section.
Also, one likely effect of a departure from chemical equilibrium
in the hadronic phase is an excess of $\rho$ mesons, increasing
the strength of the peak and
thus accounting for some of the missing signal.
The overall production rate is sensitive to $T_{fo}$:  increasing
$T_{fo}$ from 120 to 140 MeV while keeping $T_c=150$ MeV
gives $d^2N_{e^+e^-}^{(\rho^0+\omega)}
/dm d\eta = 8.5 \times 10^{-4}$ GeV$^{-1}$
($T_{fo}=140$ MeV with $T_c=200$ MeV gives
$d^2N_{e^+e^-}^{(\rho^0+\omega)}/dm d\eta =1.5 \times 10^{-3}$ GeV$^{-1}$);
on the other hand, sensitivity to increasing $\tau_0$ or decreasing
$x$ is minimal.
Omitting transverse expansion gives a value that is
a factor of 2-3
too high.
Since  the estimated error in the NA45 data
is of order 50\%, we would not rule out any of the
calculations except that with no transverse expansion.

Partial restoration of chiral symmetry in hot and dense matter \cite{ko}
is expected to modify the masses and lifetimes (widths) of vector
mesons. Changes in the $\rho$-meson properties may indicate chiral
restoration through modified $e^+e^-$ production. However, there is no
consensus at present on the behavior of the $\rho$ \cite{pisar}.
Arguments based on QCD sum rules \cite{hatsuda} and effective
Lagrangians \cite{jean} point to decreasing meson masses with increasing
temperature and density, while vector-meson dominance studies \cite{gale}
and consistency arguments \cite{ellis} have been used to support the
opposing view.
 For the time being, we therefore use the free mass and
width of the resonances as a first approximation.
The study of the effects of in-medium modifications on $e^+e^-$
production from thermal resonances as a probe for chiral restoration
is left for future work.

It is tempting to carry out a calculation without a transition to a
high-energy phase in the present model. However, this would not be
sensible with the massless pion gas approximation.
The complete resonance spectrum would have to be taken into account
and such an investigation is beyond the scope of the present paper.

We have shown that transverse expansion does not destroy the correlation
suggested in Ref.\cite{rho}.
If $T_c$ is close to the high end of the interval considered here
($T_c \simeq 200$ MeV) and $T_{fo}$ is not too low, we come close to
reproducing the preliminary NA45 data, so that the shape of the $e^+e^-$
$m_T$ spectrum, as parametrized by $T_{fit}$, should make a good thermometer
to measure $T_c$ in this region.  If, however, $T_c \simeq 150$ MeV or
$T_{fo} < 140$ MeV, the contribution from other (potentially long-lived)
resonances appears to be significant, and therefore the usefulness of the
$\rho$ meson as a thermometer in that region is questionable.
The preliminary NA45 data appear to rule out
models in which transverse expansion plays no
role, but otherwise place no significant constraints on the collision
dynamics.

\acknowledgements

We thank T. Ullrich for providing the preliminary NA45 data, and
C.~Gale for useful comments.  We also thank the Institute for Nuclear
Theory at the University of Washington for its hospitality and partial
support.  This work was supported in part by the U.S. Department of
Energy under Grant No.\ DOE/DE-FG02-86ER-40251,
in part by the Natural Sciences and Engineering
Research Council of Canada, and in part by the FCAR fund of the
Qu\'ebec government.

\newpage
\section*{Figure Captions}

\begin{description}
\item[Fig.~1:] A central S+Au collision at SPS energy, using our
standard set of parameters and $T_c$=170 MeV.
\item[Fig.~2:] $T_{fit}$ for various parameter sets as a function of
$T_c$.
\end{description}

\vfill \eject

\end{document}